\documentclass[journal]{IEEEtran}

\ifCLASSINFOpdf
   \usepackage[pdftex]{graphicx}
   \DeclareGraphicsExtensions{.pdf,.jpeg,.png}
\else
\fi
\usepackage{algorithmic}

\usepackage{float}
\usepackage{amsmath}

\usepackage{cite}
\def\BibTeX{{\rm B\kern-.05em{\sc i\kern-.025em b}\kern-.08em
    T\kern-.1667em\lower.7ex\hbox{E}\kern-.125emX}}
\begin{document}
%
\title{Partial FOV Center Imaging (PCI): \\ A Robust X-Space Image Reconstruction for Magnetic Particle Imaging}
%
%
%



\author{Semih Kurt, Yavuz Muslu, and Emine Ulku Saritas

\thanks{This work was supported by the Scientific and Technological Research Council of Turkey (Grant No: TUBITAK 115E677). (Corresponding author: Semih Kurt.)}

\thanks{S. Kurt is with the Department of Electrical and Electronics Engineering, Bilkent University, 06800 Ankara, Turkey, and also with the National Magnetic Resonance Research Center (UMRAM), Bilkent University, 06800 Ankara, Turkey (e-mail: kurt@ee.bilkent.edu.tr).}

\thanks{Y. Muslu was with the Department of Electrical and Electronics Engineering, Bilkent University, 06800 Ankara, Turkey, also with the National Magnetic Resonance Research Center (UMRAM), Bilkent University, 06800 Ankara, Turkey. He is now with the Department of Biomedical Engineering, University of Wisconsin-Madison, Madison, Wisconsin 53706, USA, and also with the Department of Radiology, University of Wisconsin-Madison, Madison, Wisconsin 53706, USA.}

\thanks{E. U. Saritas is with the Department of Electrical and Electronics Engineering, Bilkent University, 06800 Ankara, Turkey, also with the National Magnetic Resonance Research Center (UMRAM), Bilkent University, 06800 Ankara, Turkey, and also with the Neuroscience Program, Sabuncu Brain Research Center, Bilkent University, 06800 Ankara, Turkey.}

}
\maketitle

\begin{abstract}
Magnetic Particle Imaging (MPI) is an emerging medical imaging modality that images the spatial distribution of superparamagnetic iron oxide (SPIO) nanoparticles using their nonlinear response to applied magnetic fields. In standard x-space approach to MPI, the image is reconstructed by gridding the speed-compensated nanoparticle signal to the instantaneous position of the field free point (FFP). However, due to safety limits on the drive field, the field-of-view (FOV) needs to be covered by multiple relatively small partial field-of-views (pFOVs). The image of the entire FOV is then pieced together from individually processed pFOVs. These processing steps can be sensitive to non-ideal signal conditions such as harmonic interference, noise, and relaxation effects. In this work, we propose a robust x-space reconstruction technique, Partial FOV Center Imaging (PCI), with substantially simplified pFOV processing. PCI first forms a raw image of the entire FOV by mapping MPI signal directly to pFOV center locations. The corresponding MPI image is then obtained by deconvolving this raw image by a fully known, compact kernel, which solely depends on the pFOV size. We analyze the performance of the proposed reconstruction via extensive simulations, as well as imaging experiments on our in-house FFP MPI scanner. The results show that PCI offers a trade-off between noise robustness and interference robustness, outperforming standard x-space reconstruction in terms of both robustness against non-ideal signal conditions and image quality. 
\end{abstract}


%
\IEEEpeerreviewmaketitle

\section{Introduction}

\IEEEPARstart{M}{agnetic} Particle Imaging (MPI) leverages the nonlinear magnetization response of superparamagnetic iron oxide (SPIO) nanoparticles, to generate an image of their spatial distribution with high resolution, contrast, and sensitivity \cite{Gleich2005,Weizenecker_2007,X-spaceSafe,SARITAS2013116,Bauer2015,Zheng2017}. Three different magnetic fields are utilized to obtain the MPI signal. A static selection field with a strong gradient creates a field free point (FFP). A sinusodial drive field excites the nanoparticles in the vicinity of the FFP, effectively scanning a field-of-view (FOV) via moving the FFP. However, safety limits restrict the size of the FOV that can be scanned by the drive field alone to a few cm or less \cite{6510534,schmaleSafety}. The small FOV covered by the drive field is called a partial FOV (pFOV) \cite{6494648} or a patch \cite{patch1,patch2,patch3}. Then, to cover larger FOVs,  low-frequency focus fields are employed \cite{slew-rate}. Due to the limits on the slew rates of the focus fields \cite{slewlimit,schmaleSafety}, a realistic scan trajectory may consist of numerous highly-overlapping small pFOVs. In x-space reconstruction, these pFOVs are first individually processed by gridding the speed-compensated signal to the trajectory of the FFP, and then combined to form the image of the entire FOV \cite{5482192,5728922,6494648,convex_x_space,alper,pulsedMPI}. The resulting MPI image shows the spatial distribution of the nanoparticles blurred by the point spread function (PSF) of the imaging system.

\par One of the challenges in MPI is the direct feedthrough signal caused by the simultaneous excitation and reception, as it can be many orders of magnitude larger than the nanoparticle signal \cite{5482192,5728922}. To keep the direct feedthrough contained to the fundamental harmonic, the drive field is typically chosen as a pure sinusoid. Then, a gradiometric receive coil \cite{7107082,ProjectionXspace,Analogreceivesignalprocessing}, active/passive compensation \cite{ActivePassiveCompensation}, and/or analog/digital filtering can be utilized to counteract the effects of the direct feedthrough. During image reconstruction with the x-space approach, the fundamental harmonic lost due to filtering is recovered via enforcing smoothness and non-negativity on the reconstructed image \cite{6494648,gazimmfd337864}. If the received signal also contains higher harmonics due to system non-idealities and/or interferences, however, simple filtering no longer suffices. In practice, therefore, a background measurement is subtracted from the received signal to cancel out potential higher harmonic interferences. If the interference level is comparable to the nanoparticle signal, or if there is a drift in the system, this background cancellation may not work as desired. A detailed analysis on the effects of higher harmonic interference on the quality of the reconstructed MPI image has not yet been presented.

\par In this work, we present an x-space image reconstruction technique called "pFOV center imaging" (PCI), which features substantially simplified pFOV processing and increased robustness against harmonic interferences. The proposed technique first forms a raw image of the entire FOV by mapping the MPI signal directly to the pFOV center locations. Then, this raw image is deconvolved by a fully known, compact kernel to obtain the corresponding MPI image. Importantly, this kernel solely depends on the pFOV size, and is independent of the other scanning parameters or the nanoparticle type. We analyze the performance of the proposed method at different signal-to-noise ratio (SNR) and harmonic interference levels, demonstrating a trade-off between noise robustness and harmonic interference robustness. With extensive simulations, as well as imaging experiments on a FFP scanner, we show that PCI outperforms standard x-space reconstruction in terms of image quality, noise robustness, and interference robustness.

\section{Theory}
For a trajectory that contains a drive field superimposed with a low-frequency focus field, the time-domain MPI signal, $s\left(t\right)$, can be written as \cite{5482192,5728922}:
\begin{equation}
\label{eqn:time-domain MPI signal}
 s\left(t\right) = \alpha \dot{x}_s\left(t\right)  \hat{\rho}\big(x_s\left(t\right)\big)
\end{equation}
where 
\begin{equation}
\label{eqn:ideal MPI image}
 \hat{\rho}\big(x_s\left(t\right)\big) = \rho\left(x\right) \ast  h\left(x \right) \biggl|_{x=x_s\left(t\right)}
\end{equation}
Here, $x_s\left(t\right)$ is the instantaneous FFP position, $\dot{x}_s\left(t\right)$ is the instantaneous FFP speed, $\rho (x)$ is the particle distribution, $h\left(x \right)$ is the PSF, and $\hat{\rho} (x)$ is the PSF-blurred "ideal" MPI image. In addition, $ \alpha $ is a constant that depends on the selection field gradient, the nanoparticle type, the magnetic moment of the nanoparticle, and the sensitivity of the receive coil \cite{5482192,5728922}. For the following derivations, we ignore the nanoparticle relaxation effects on the signal.

\par Due to direct feedthrough filtering, the MPI signal loses its fundamental harmonic component. It has been shown that the contribution of the lost first harmonic for each pFOV is a DC term \cite{rahmersignalencoding,6494648}. In standard x-space reconstruction, DC terms are recovered via pFOV stitching by enforcing non-negativity and continuity constraints on the reconstructed image \cite{6494648}.

\subsection{Partial FOV Center Imaging (PCI):}
Let $x_{0j}$ be the center position of the $j^{th}$ pFOV and $t_{0j}$ be the time instant when the FFP is at $x_{0j}$, i.e.,  
\begin{equation}
\label{eqn:center of pFOV}
 x_s\left(t\right) \biggl|_{t=t_{0j}}  = x_{0j}  \; \; \; ,\:\text{for}\: j=1,...,N
\end{equation}
where $N$ is the total number of pFOVs. We propose to sample $s\left(t\right)$ at the centers of pFOVs to get a raw image $\hat{\rho}_{0} (x)$ such that
\begin{subequations}
\begin{align} 
 \hat{\rho}_{0}\left(x_{0j}\right) &= s\left(t\right) \biggl|_{t=t_{0j}} \\
\label{eqn:RPCI as sampled s(t)}
&= \alpha \dot{x}_s\left(t_{0j}\right)  \hat{\rho}\left(x_{0j}\right)\\
\label{eqn:RPCI as sampled s(t) final}
&= \beta_0  \hat{\rho}\left(x_{0j}\right) \; \; \; ,\:\text{for}\: j=1,...,N
\end{align}
\end{subequations}
For the linear trajectories considered in this work, the FFP speed is identical at the centers of pFOVs. Therefore $\beta_0=\alpha \dot{x}_s\left(t_{0j}\right)$ is a constant for $j=1,...,N$. 

\par In practice, due to direct feedthrough filtering, $\hat{\rho}_{0}\left(x\right) $ is devoid of the contribution of the first harmonic. We have previously shown that the lost DC term for the $j^{th}$ pFOV can be represented by a convolution as \cite{7960614}:
\begin{equation}
\label{eqn:dc image}
 \hat{\rho}_{dc}\left(x_{0j}\right) = \frac{4}{\pi W} \left(\hat{\rho}\left(x\right) \ast \sqrt{1-\left(\frac{2x}{W}\right)^2}\right) \biggl|_{x=x_{0j}} 
\end{equation}
where $W$ is the extent of each pFOV. Taking into account the lost DC term and using Eqs. \ref{eqn:RPCI as sampled s(t) final} and \ref{eqn:dc image}, the raw image at $x_{0j}$ can be expressed as
\begin{subequations}
\begin{equation}
\label{eqn:RPCI with filtering}
  \Tilde{\rho}_{0}\left(x_{0j}\right)  = \beta_0  \big(\hat{\rho}\left(x_{0j}\right)-\hat{\rho}_{dc}\left(x_{0j}\right)\big)
\end{equation}
\begin{equation}
\label{eqn:RPCI explicit}
=    \beta_0 \left( \hat{\rho}\left(x\right)\ast \left( \delta \left(x\right)-\frac{4}{\pi W} \sqrt{1-\left(\frac{2x}{W}\right)^2}\right) \right)\biggl|_{x=x_{0j}} 
\end{equation}
\end{subequations}
valid for all $j=1,...,N$. Hence, the raw image $\Tilde{\rho}_{0}\left(x\right)$ can be written as the ideal image convolved with a kernel, i.e.,
\begin{equation}
\label{eqn:RPCI final}
 {\Tilde{\rho}}_{0}\left(x\right) = \hat{\rho}\left(x\right)\ast h_{0}\left(x\right)
\end{equation}
where 
\begin{equation}
\label{eqn:PCI kernel}
 h_{0}\left(x\right) = \beta_0 \left( \delta \left(x\right)-\frac{4}{\pi W} \sqrt{1-\left(\frac{2x}{W}\right)^2}\right)
\end{equation}
Note that $h_{0} (x)$ is a fully known, compact kernel with full-width $W$. Importantly, it does not depend on the nanoparticle type. Next, we can deconvolve $\Tilde{\rho}_{0}\left(x\right)$ by the known kernel $h_{0} (x)$ to obtain $\hat{\rho} (x)$, i.e.,
\begin{equation}
\label{eqn:PCI deconv}
 \hat{\rho}\left(x\right)  = \Tilde{\rho}_{0}\left(x\right) \ast^{-1} h_{0}\left(x\right)
\end{equation}
Here, $\ast^{-1}$ denotes the deconvolution operation. We refer to this technique as pFOV center imaging (PCI).

\begin{figure}[h]
\centering
\includegraphics{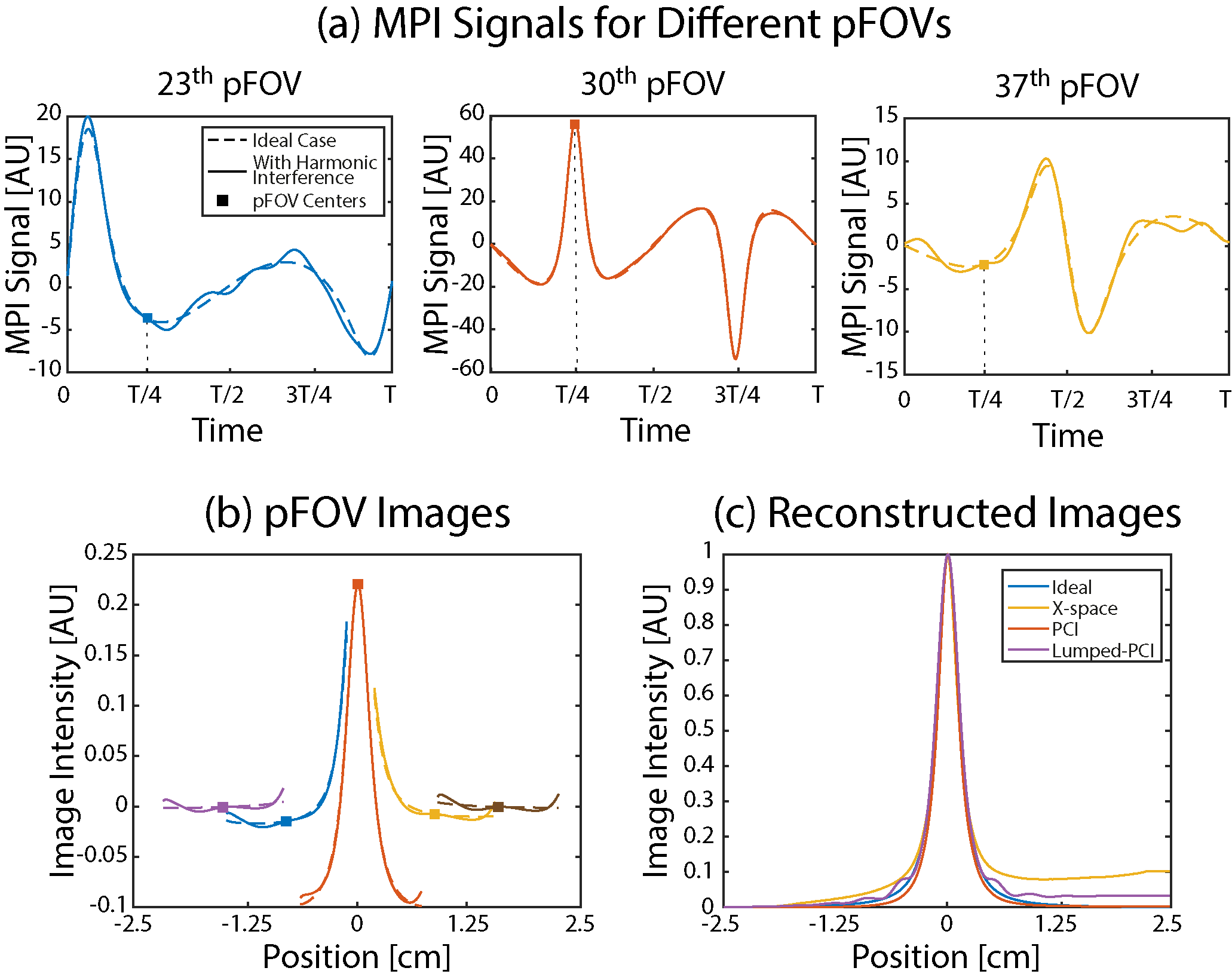}
\caption{Effects of harmonic interference on the MPI signal and image reconstruction. (a) The signals after direct feedthrough filtering and (b) the reconstructed images from selected pFOVs with and without harmonic interference. The centers of pFOVs are particularly robust against interferences, while the edges can exhibit large deviations. (c) The reconstructed MPI images for standard x-space reconstruction, PCI, and Lumped-PCI. Standard x-space reconstruction exhibits an accumulated error toward one end of the image. PCI image is free from such artifacts, while Lumped-PCI shows slight ripple-like artifacts. Here, the ideal case refers to the convolution of the nanoparticle distribution with the imaging PSF. These simulations were performed for a point source placed at the origin, with 2.4 T/m/${\mu}_0$ gradient, 10 mT drive field at 9.7 kHz, 10 T/s slew-rate.}
\label{fig: motivation}
\end{figure}

\par Because PCI utilizes only the signals at pFOV centers, it is particularly robust against harmonic interference effects. This robustness is directly related to the MPI harmonic image basis set, which is composed of Chebyshev polynomials of the second kind \cite{rahmersignalencoding,6494648}. Independent of the harmonic number, each image basis varies relatively slowly near the pFOV centers when compared to locations towards the edges. In return, the pFOV centers are particularly robust against harmonic interference effects, whereas the edges are sensitive to such interferences. A pictorial depiction of this effect is demonstrated in Fig. \ref{fig: motivation}, where the MPI signal and corresponding images from different pFOVs are shown with and without harmonic interference (see Section \ref{Noise and Interference} for details on these simulations). As seen Fig. \ref{fig: motivation}a-b, the MPI signals and pFOV images corresponding to the pFOV centers are approximately the same with and without interference, whereas other positions can exhibit large deviations. These deviations make it more difficult to estimate the lost DC term, causing standard x-space reconstruction to exhibit an accumulated error toward one end of the image, as shown in Fig. \ref{fig: motivation}c. On the other hand, the MPI image reconstructed using PCI does not exhibit such an artifact. 

\subsection{Lumped PCI:}
Since PCI uses only a small portion of the received signal, it may be affected by noise. To increase robustness against noise, PCI can be modified to use all the received signal. Let $x_{kj}$ be the $k^{th}$ position in the $j^{th}$ pFOV and $t_{kj}$ be the time instant when the FFP is at $x_{kj}$, i.e., 
\begin{align}
\label{eqn:position k of pFOVs}
 x_s\left(t\right) \biggl|_{t=t_{kj}}  = x_{kj}  \; \; \; ,\:\text{for}\: &j=1,...,N, \\
 \; \; \; \:\: &k=-K,...,K \nonumber
\end{align}
where $2K+1$ is the number of samples in one half drive field cycle. A raw image $\hat{\rho}_{k}\left(x\right)$ can be constructed by sampling $s(t)$ at positions $x_{kj}$, but assigning these samples to center positions $x_{0j}$:
\begin{subequations}
\begin{align}
\label{eqn:Raw image k without filter}
 \hat{\rho}_{k}\left(x_{0j}\right) &= \alpha \dot{x}_s\left(t_{kj}\right)  \hat{\rho}\left(x_{kj}\right) \\
 &= \beta_k \hat{\rho}\left(x_{kj}\right)
\end{align} 
\end{subequations}
Note that the FFP speed is identical at the  $k^{th}$ positions of pFOVs. Therefore $\beta_k= \alpha \dot{x}_s\left(t_{kj}\right)$ is a constant for $j=1,...,N$. 

\par Direct feedthrough filtering causes the same DC loss at all positions in a given FOV \cite{6494648,7960614}. Hence, similar to Eq. \ref{eqn:RPCI with filtering}, the raw image with lost DC term can be expressed as
\begin{equation}
\label{eqn:Raw image k with filtering p1}
 \Tilde{\rho}_{k}\left(x_{0j}\right)  = \beta_k  \big(\hat{\rho}\left(x_{kj}\right)-\hat{\rho}_{dc}\left(x_{0j}\right)\big)
\end{equation}
valid for all $j=1,...,N$. By comparing Eqs. \ref{eqn:RPCI with filtering} and \ref{eqn:Raw image k with filtering p1},
\begin{subequations}
\begin{equation}
\label{eqn:Raw image k with filtering p2}
 \Tilde{\rho}_{k}\left(x_{0j}\right)  = \beta_k  \Big(\hat{\rho}\left(x\right) \ast \delta \big(x-(x_{0j}-x_{kj})\big)\biggl|_{x=x_{0j}} -\hat{\rho}_{dc}\left(x_{0j}\right)\Big)
\end{equation}
\begin{equation*}
\label{eqn:Raw image k explicit2 part1}
= \beta_k  \Bigg( \hat{\rho}\left(x\right)\ast \bigg( \delta \left(x-(x_{0j}-x_{kj})\right)
\end{equation*}
\begin{equation}
\label{eqn:Raw image k explicit2 part2}
-\frac{4}{\pi W} \sqrt{1-\left(\frac{2x}{W}\right)^2}\bigg) \Bigg)\biggl|_{x=x_{0j}}
\end{equation}
\end{subequations}
We can rewrite the $k^{th}$ raw image $\Tilde{\rho}_{k}\left(x\right)$ in a simplified form:
\begin{equation}
\label{eqn:Raw image final}
 \Tilde{\rho}_{k}\left(x\right) = \hat{\rho}\left(x\right)\ast h_{k}\left(x\right)
\end{equation}
where
\begin{equation}
\label{eqn:Raw kernel}
 h_{k}\left(x\right) = \beta_k \left( \delta \left(x-(x_{0j}-x_{kj})\right)-\frac{4}{\pi W} \sqrt{1-\left(\frac{2x}{W}\right)^2}\right)
\end{equation}
Then, one can reconstruct $\hat{\rho} (x)$ via deconvolving $\Tilde{\rho}_{k}\left(x\right)$ by $ h_{k}\left(x\right)$:
\begin{equation}
\label{eqn:Raw deconv}
 \hat{\rho}\left(x\right)  = \Tilde{\rho}_{k}\left(x\right) \ast^{-1}  h_{k}\left(x\right)
\end{equation}
\par Eq. \ref{eqn:Raw image final} shows that sampling $s(t)$ at different positions in pFOVs can create different raw images, and one can obtain $\hat{\rho}\left(x\right)$ from any of these raw images.
Note that among all $\Tilde{\rho}_{k}\left(x\right)$, the one with the highest SNR is $\Tilde{\rho}_{0}\left(x\right)$, as the FFP speed is maximum when passing through the pFOV center. To boost the SNR and use the entire signal, we can sum all $\Tilde{\rho}_{k}\left(x\right)$ to get a raw lumped image, i.e.,
\begin{subequations}
\begin{align}
\label{eqn:RPLI image}
 \Tilde{\rho}_{lum}\left(x\right) &=\sum_{k=-K}^{K} \Tilde{\rho}_{k}\left(x\right)\\
 &= \hat{\rho}\left(x\right)\ast h_{lum}\left(x\right)
\end{align}
\end{subequations}
where
\begin{equation}
\label{eqn:PLI kernel}
 h_{lum}\left(x\right) = \sum_{k=-K}^{K} h_{k}\left(x\right)
\end{equation}
Eq. \ref{eqn:PLI kernel} follows from the linearity of the convolution operation. Once again, we can reconstruct $\hat{\rho}(x)$ via deconvolution:
\begin{equation}
\label{eqn:PLI deconv}
 \hat{\rho}\left(x\right)  = \Tilde{\rho}_{lum}\left(x\right) \ast^{-1}   h_{lum}\left(x\right)
\end{equation}
We refer to this extension of the method as Lumped-PCI.

\par While Lumped-PCI boosts SNR and improves noise robustness, it is slightly more sensitive against harmonic interferences when compared to PCI. Figure \ref{fig: motivation}c shows an example Lumped-PCI image under harmonic interference, where ripple-like artifacts can be seen. Still, Lumped-PCI displays increased reconstruction fidelity and interference robustness when compared to standard x-space reconstruction. As Lumped-PCI sums the raw images from the entire pFOV, it effectively averages out the negative/positive deviations from the ideal values.

\section{Methods}
\subsection{Simulations}\label{Simulations}
MPI simulations were carried out using a custom toolbox developed in MATLAB (Mathworks, Natick, MA). Simulation parameters were chosen to match the parameters of our in-house FFP MPI scanner (Fig. \ref{fig: scanner}a). Accordingly, the selection field gradients were (-4.8, 2.4, 2.4) T/m/${\mu}_0$ in (x, y, z) directions. 10 mT drive field at 9.7 kHz along the z-direction was simulated together with a focus field that creates 1 T/s slew rate in the z-direction. We used a 5$\times$5 cm$^2$ vasculature phantom and assumed 25 nm nanoparticle diameter. The pFOV size was 8.33 mm and a 2D FOV of 5$\times$5 cm$^2$ was scanned using a linear trajectory, similar to the one shown in Fig. \ref{fig: scanner}b. The entire FOV was scanned in 51 lines. To match the conditions of the imaging experiments, the simulated MPI signal was sampled at 2 MS/s.

\subsection{Noise and Harmonic Interference Robustness Analysis} \label{Noise and Interference}
\par To analyze the robustness of the proposed method, we simulated noise, harmonic interference, and relaxation effects on MPI signal. For noise analysis, white Gaussian noise was added to the time-domain MPI signal at 10 different noise levels, with signal-to-noise ratio (SNR) varying between 5-50~dB. SNR was defined using the peak signal amplitude as follows:
\begin{equation}
\label{eqn:SNR}
SNR  = 20 log_{10} \Bigg(\frac{\max \limits_{t} \big|s\left(t\right)\big|}{\sigma} \Bigg)
\end{equation}
Here, $\sigma$ denotes the standard deviation of noise, and $s\left(t\right)$ is the MPI signal after direct feedthrough filtering.

\par For harmonic interference analysis, harmonic interference was added to the spectrum of $s\left(t\right)$. When the drive field is applied alone, the spectrum of the MPI signal contains only the harmonics of the fundamental frequency, $f_0$  \cite{5482192}. However, for the linear scan trajectory used in this work, the harmonics spread to very narrow nearby bands \cite{HDX-IWMPI}. Let $S(f)$ denote the Fourier transform of $s(t)$ and $S_n(f)$ denote the $n^{th}$ harmonic band, i.e.,
\begin{equation}
\label{eqn: harmonic band}
S_n(f) =
  \begin{cases}
                   S(f) & ,\; \left(n-\frac{1}{2}\right) f_0 < f < \left(n+\frac{1}{2}\right) f_0    \\
                   0 & ,\; \text{otherwise} \\
  \end{cases}
\end{equation}
\par The magnitude of the interference added to the  $n^{th}$ harmonic (i.e., at $f=n \,f_0$) was uniformly distributed between 0 and $\gamma_n$, whereas the phase of it was uniformly distributed between 0 and 2$\pi$. To assess the strength of the MPI signal against harmonic interference, we used the signal-to-interference ratio (SIR) metric. Based on our experimental observations, harmonic interference was simulated so that each harmonic band had the same SIR level, i.e.,
\begin{equation}
\label{eqn:SIR}
SIR  = 20 log_{10} \Bigg(\frac{\max \limits_{f } \big| S_n\left(f\right) \big|}{\gamma_n} \Bigg)
\end{equation}
Hence, as the magnitude spectrum in MPI decayed at higher harmonics, the interference followed the same trend. For simulations, 6 different SIR levels between 4-24 dB were tested.

\par First, we simulated noise and harmonic interference effects separately. Then, we incorporated both effects simultaneously with SNR ranging between 5-50 dB and SIR ranging between 4-24 dB. Monte Carlo simulations were performed via repeating each case 50 times. Next, to incorporate the effects of relaxation, we utilized a realistic time constant of $\tau$ $=$ $3$ $\mu s$ \cite{8322193,mustafaviscomakale}, using the model provided in \cite{6297476}. For this analysis, SNR was fixed at 30 dB and SIR at 8 dB. 

\subsection{Imaging Experiments}\label{ImagingExp}
\begin{figure}[h]
\centering
\includegraphics{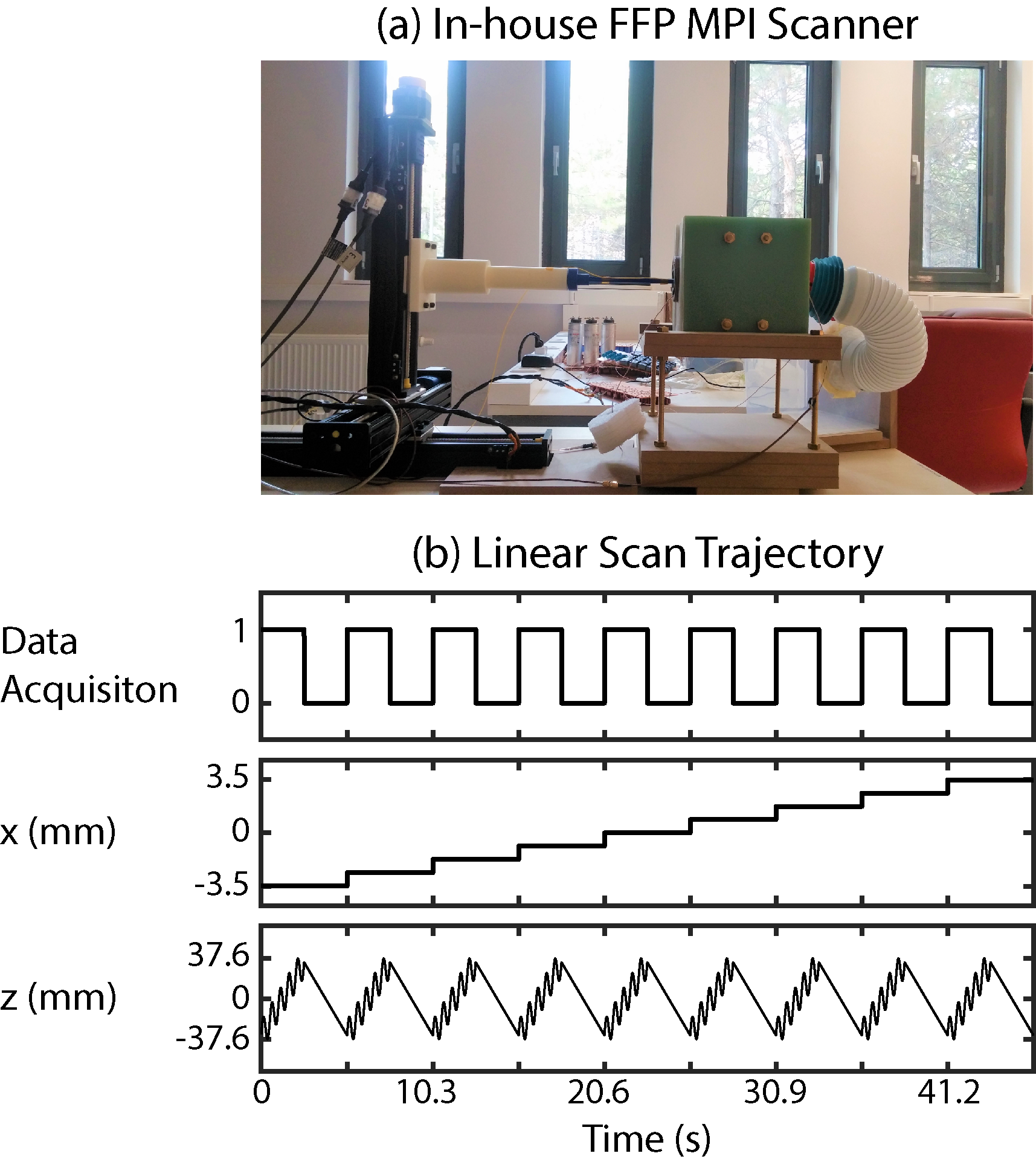}
\caption{An overview of our in-house FFP MPI scanner and the linear scan trajectory used in the imaging experiments. (a) This scanner features (-4.8, 2.4, 2.4) T/m/${\mu}_0$ selection field gradients in (x, y, z) directions, with a maximum FOV of 1$\times$1$\times$10 cm$^3$. (b) The linear scan trajectory had 0.7$\times$7.52 cm$^2$  FOV in x-z plane, 0.07 T/s slew rate along z-direction using continuous robotic arm motion, with 23.2 sec active scan time. This figure shows a simplified version of the actual trajectory: instead of the drive field at 9.7 kHz and 10 mT-peak, a representative drive field at a much lower frequency is plotted.}
\label{fig: scanner}
\end{figure}

Imaging experiments were performed on in-house FFP MPI scanner (Fig. \ref{fig: scanner}) \cite{8479214}. The selection field of this scanner was generated by two permanent magnets with 7-cm diameter and 2-cm thickness, placed at 8-cm separation. The resulting selection field gradients were  (-4.8, 2.4, 2.4) T/m/${\mu}_0$ in (x, y, z) directions. The drive field coil had a 1.5 mT/A sensitivity, with 95\% homogeneity in a 4.5-cm long region, and was built using 3 layers of Litz wire with 80 turns. For the receive coil, a three-section gradiometer type coil with 34 and 17.5 windings for the main section and side sections was utilized \cite{7107082}. The drive and receive coils were positioned coaxially, and placed inside a cylindrical copper shield with 1-cm thickness at the center of the magnet configuration. The maximum FOV of this FFP MPI scanner is 1$\times$1$\times$10 cm$^3$. 


\par In imaging experiments, 0.7x7.52 cm$^2$ FOV in x-z plane was scanned. To cover this FOV, the linear trajectory shown in Fig. \ref{fig: scanner}b was utilized. The drive field was at 9.7 kHz and 10 mT-peak along the z-direction, resulting in a 8.33 mm pFOV length. Instead of a focus field, a three-axis robotic arm (Motor-Driven Velmex BiSlide, Model: MN10-0100-E01-21) was used to move the phantom continuously along the z-direction, and stepwise along the x-direction. The maximum slew rate for the continuous motion was 0.07 T/s due to the speed limitation of this mechanical system. The entire FOV was scanned in 9 lines, with an active scan time of 23.2 sec.


\subsection{Imaging Phantoms}
\par Two different imaging phantoms were prepared to demonstrate the performance of the proposed method. For the first phantom, two 3-mm vials were filled with Perimag (Micromod GmbH, Germany) nanoparticles with a diluted concentration of 5 mg Fe/mL. Vials were separated by 9-mm distance along the z-direction. 

\par To show that the proposed method can successfully handle different nanoparticle types, a second phantom was prepared. For this phantom, three 2-mm inner diameter vials were filled with Nanomag-MIP (Micromod GmbH, Germany) nanoparticles with a diluted concentration of 1.43 mg Fe/mL, Vivotrax (Magnetic Insight Inc., USA) with an undiluted concentration of 5.5 mg Fe/mL, and a homogeneous mixture of the two. The vials were separated by 15-mm distances along the z-direction.

\subsection{Signal Pre-processing \& Image Reconstruction} \label{Signal Pre-Processing}
We used the same signal pre-processing steps in both the simulations and experiments. First, the received signal was digitally high-pass filtered to remove any remaining direct feedthrough of the drive field. Next, for only experiments, a low-pass filter was applied to filter out the signal near and after the self-resonance frequency of the receive coil, which was measured at around 280~kHz. The resulting signal was further filtered for the purposes of spectrum cleaning. For the linear trajectories used in this work, we defined pass-bands around the harmonics, with a bandwidth of 420~Hz for simulations. Due to the relatively slower slew rate in the experiments, the signal spread to a narrow band around the harmonics. Therefore, a smaller pass-band bandwidth of 16~Hz was utilized for the experiments.

\par For PCI and Lumped-PCI, the pre-processed signal values were directly used to form the raw images, as described in Eq. \ref{eqn:Raw image k without filter}. Then, these images were interpolated to a finer grid of 0.05-mm spacing and deconvolved by the kernels computed from Eq. \ref{eqn:Raw kernel}  using the built-in \textit{deconvreg} function in MATLAB, to reconstruct the final MPI images.

\subsection{Comparison of Image Quality}
Standard x-space reconstruction with DC recovery algorithm \cite{6494648} and SNR optimized pFOV stitching \cite{gazimmfd337864} was implemented for comparison purposes. The comparison and proposed techniques used identical signal pre-processing steps, as described in Section \ref{Signal Pre-Processing}.

\par For quantitative image quality assessment of the reconstructed images in simulations, the PSNR metric was employed:
\begin{equation}
\label{eqn:PSNR}
PSNR \left(I\right) = 10 log_{10} \big(\frac{R^2}{MSE} \big)
\end{equation}
where 
\begin{equation}
\label{eqn:MSE}
MSE = \frac{\sum_{M,N} \big( I[m,n]-I_{ref}[m,n] \big) ^2 }{MN}
\end{equation}
Here, $I[m,n]$ is the reconstructed image, R is the peak value that a pixel can have, $I_{ref} [m,n]$ is the reference image, MSE is the mean-squared-error between the reconstructed image and the reference image, and images are of size $M \times N$. The phantom itself was used as the reference image, and all reconstructed images were individually normalized to [0~1] range. Accordingly, higher PSNR values indicate higher fidelity image reconstruction.

\section{Results}
\subsection{Simulation Results} 
\begin{figure}[h]
\centering
\includegraphics{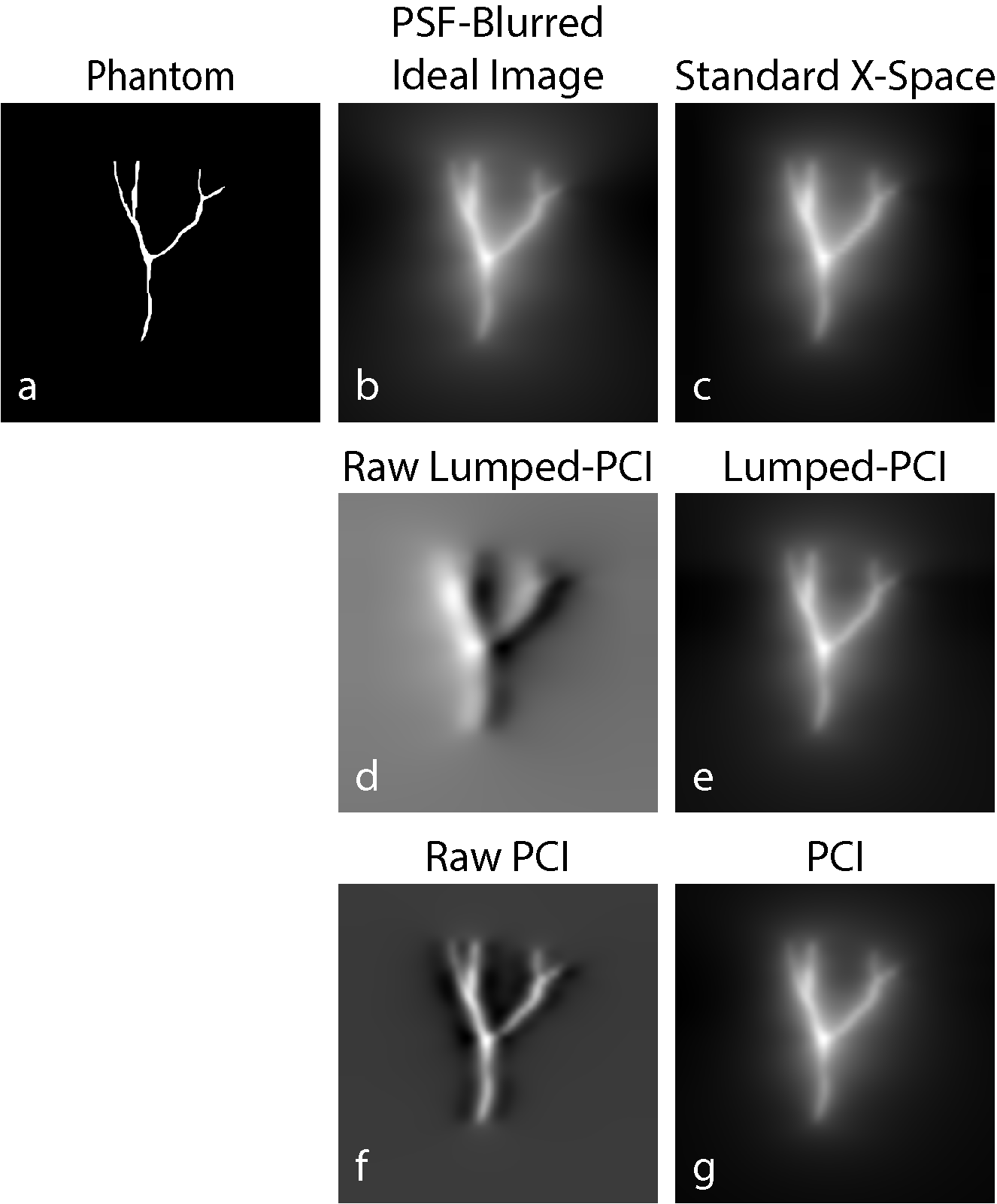}
\caption{Simulation results for ideal signal conditions. (a) A 5$\times$5 cm$^2$ vasculature phantom and (b) the corresponding PSF-blurred ideal MPI image. Images reconstructed using (c) standard x-space reconstruction, (e) Lumped-PCI, and (g) PCI. Here, (d) and (f) show the raw images for Lumped-PCI and PCI, respectively. Under ideal signal conditions, all three methods provide visually similar results.}
\label{fig: ideal simulation}
\end{figure}

Figure \ref{fig: ideal simulation} shows the results of the proposed and comparison reconstructions for a 5$\times$5 cm$^2$ vasculature phantom for the ideal case, where noise, interference, and relaxation effects are neglected. Figure 4a-4c show the phantom, the PSF-blurred ideal MPI image, and the corresponding standard x-space reconstruction result. Figure 4d-4g show the results of Lumped-PCI and PCI reconstructions together with the corresponding raw images. For this ideal case, all three methods yield visually similar results. A quantitative comparison of these images yields 13.4 dB, 13.5 dB, and 14.1 dB PSNR values for standard x-space, Lumped-PCI, and PCI, respectively. Although there is no visible difference between the reconstructed images, the PSNR values suggest that the PCI method is the most successful reconstruction method under ideal signal conditions.  
\par
\begin{figure}[h]
\centering
\includegraphics{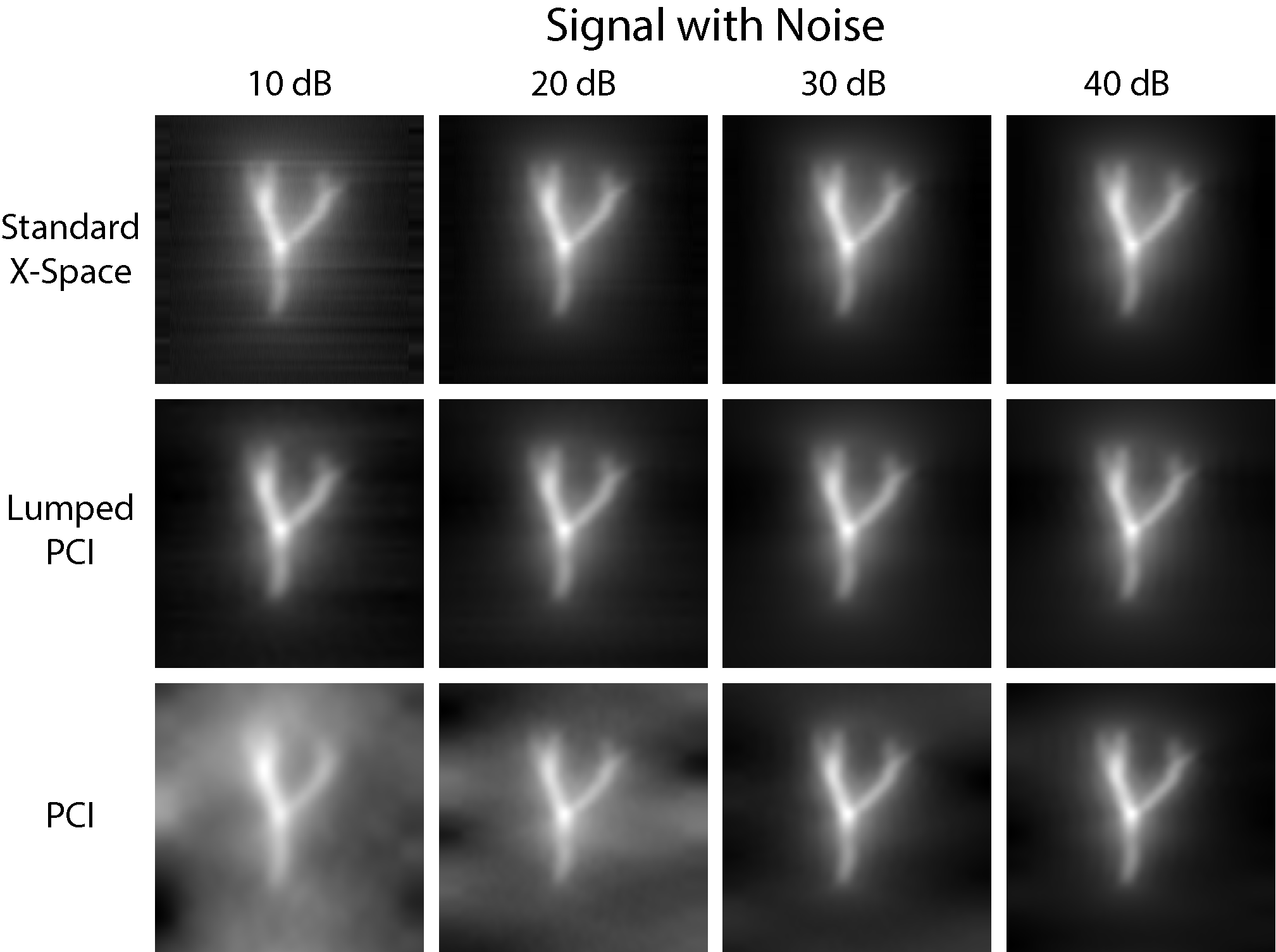}
\caption{Example results at 4 different SNR levels between 10-40 dB. Both standard x-space and Lumped-PCI methods display robustness against noise, with slight horizontal stripe artifacts in standard x-space and an overall blurring in Lumped-PCI at very low SNR levels. PCI shows degradation in image quality for the lowest two SNR levels, as it uses only a small portion of the received signal.}
\label{fig: 88 noise images}
\end{figure}

 Figure \ref{fig: 88 noise images} shows example results for the proposed and comparison reconstructions at 4 different SNR levels between 10-40 dB. Both the standard x-space and Lumped-PCI methods display robustness against noise. For standard x-space, horizontal stripe artifacts appear for the lowest SNR level of 10 dB, stemming from the inconsistencies during the stitching operation for different lines in the image. In contrast, an overall blurring is visible at the same SNR level for Lumped-PCI, due to the trade-off between noise regularization and resolution. The PCI method, on the other hand, shows degradation in image quality for the lowest two SNR levels. This relative noise sensitivity is expected, as PCI uses only a small portion of the received signal.

\par
\begin{figure}[h]
\centering
\includegraphics{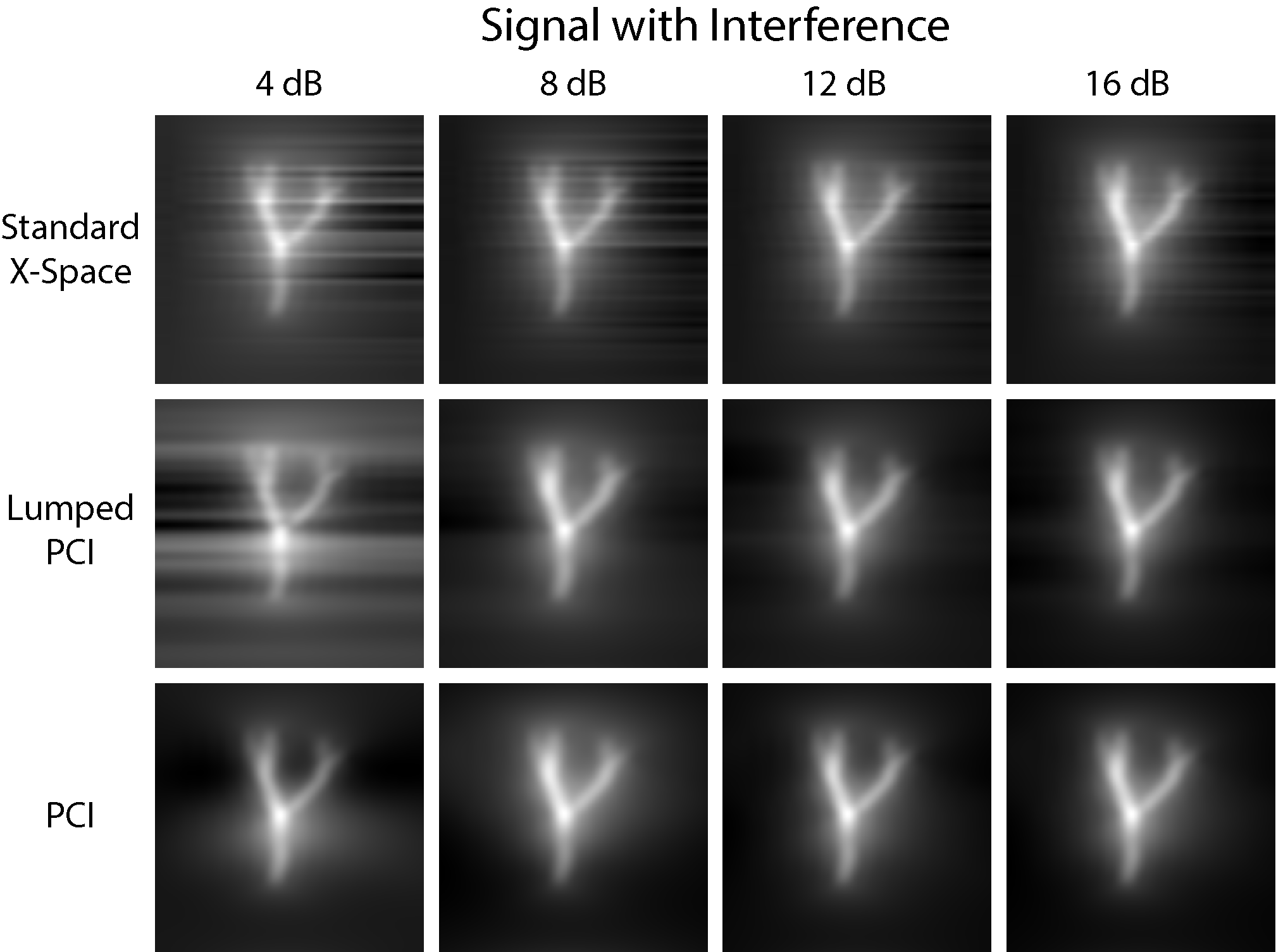}
\caption{Example results at 4 different SIR levels between 4-16 dB. Standard x-space reconstruction suffers from horizontal stripe artifacts that are visible even at the highest SIR level of 16 dB. Lumped-PCI is more robust against interference effects, however, thicker horizontal stripe artifacts arise at the lowest SIR level of 8 dB. PCI demonstrates robustness against interference at all SIR levels.}
\label{fig: 88 int images}
\end{figure}

 Figure \ref{fig: 88 int images} shows example results at 4 different SIR levels between 4-16 dB. Standard x-space reconstruction displays sensitivity against harmonic interference, manifested as horizontal stripe artifacts that are visible even at the highest SIR level of 16 dB. While Lumped-PCI is considerably more robust against interference effects, similar but thicker horizontal stripes emerge at the lowest SIR level of 8 dB. In contrast, PCI demonstrates robustness against interference at all SIR levels tested. At the lowest SIR level, while there are image intensity differences in the background, these low-resolution features do not hinder the delineation of the phantom. 

\par
\begin{figure}[h]
\centering
\includegraphics{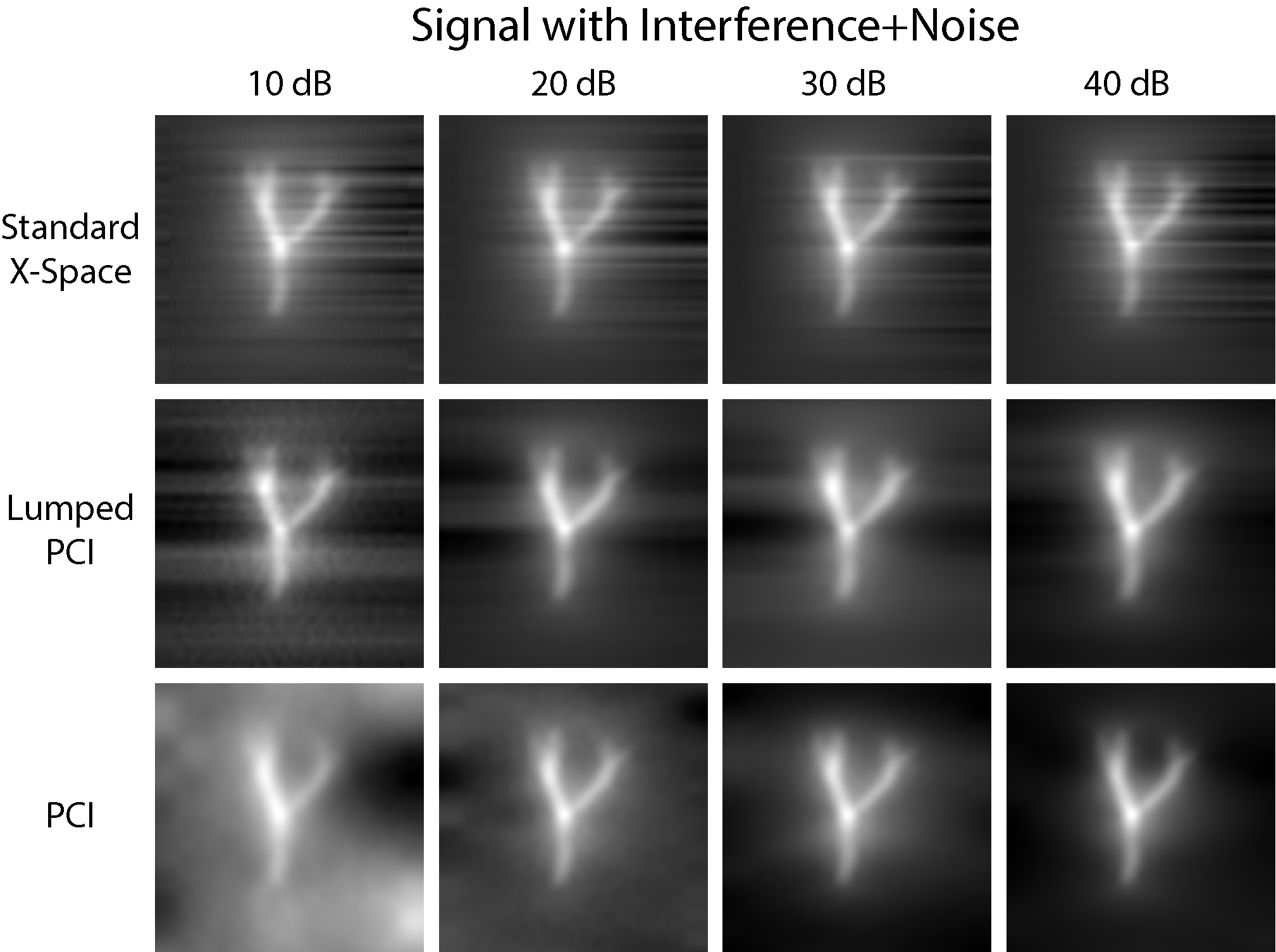}
\caption{Example results at 4 different SNR levels between 10-40 dB, with SIR fixed at 8 dB. Due to harmonic interference, the standard x-space method exhibits stripe artifacts even at the highest SNR level of 40 dB. Lumped-PCI shows improved image quality at all SNR levels, with relatively better behaved artifacts. PCI shows almost no artifacts at the highest two SNR levels, whereas its image quality degrades for the lowest two SNR levels. }
\label{fig: 88 noise+interference}
\end{figure}

 Figure \ref{fig: 88 noise+interference} displays the combined effects of noise and harmonic interference on the three reconstruction methods, at a fixed SIR level of 8 dB with SNR ranging between 10-40 dB. The presence of harmonic interference limits the performance of the standard x-space method, which suffers from stripe artifacts even at the highest SNR level of 40 dB. Lumped-PCI shows improved image quality at that SNR level, with relatively better behaved artifacts at lower SNR levels. In contrast, PCI shows almost no artifacts at the highest two SNR levels. However, its image quality degrades considerably for the lowest two SNR levels. These results are in line with those in Fig. \ref{fig: 88 noise images} and Fig. \ref{fig: 88 int images}. When noise is the dominant effect, Lumped-PCI performs the best, whereas when interference dominates over noise, PCI yields the highest image quality. 

\par
\begin{figure}[h]
\centering
\includegraphics[scale=0.8]{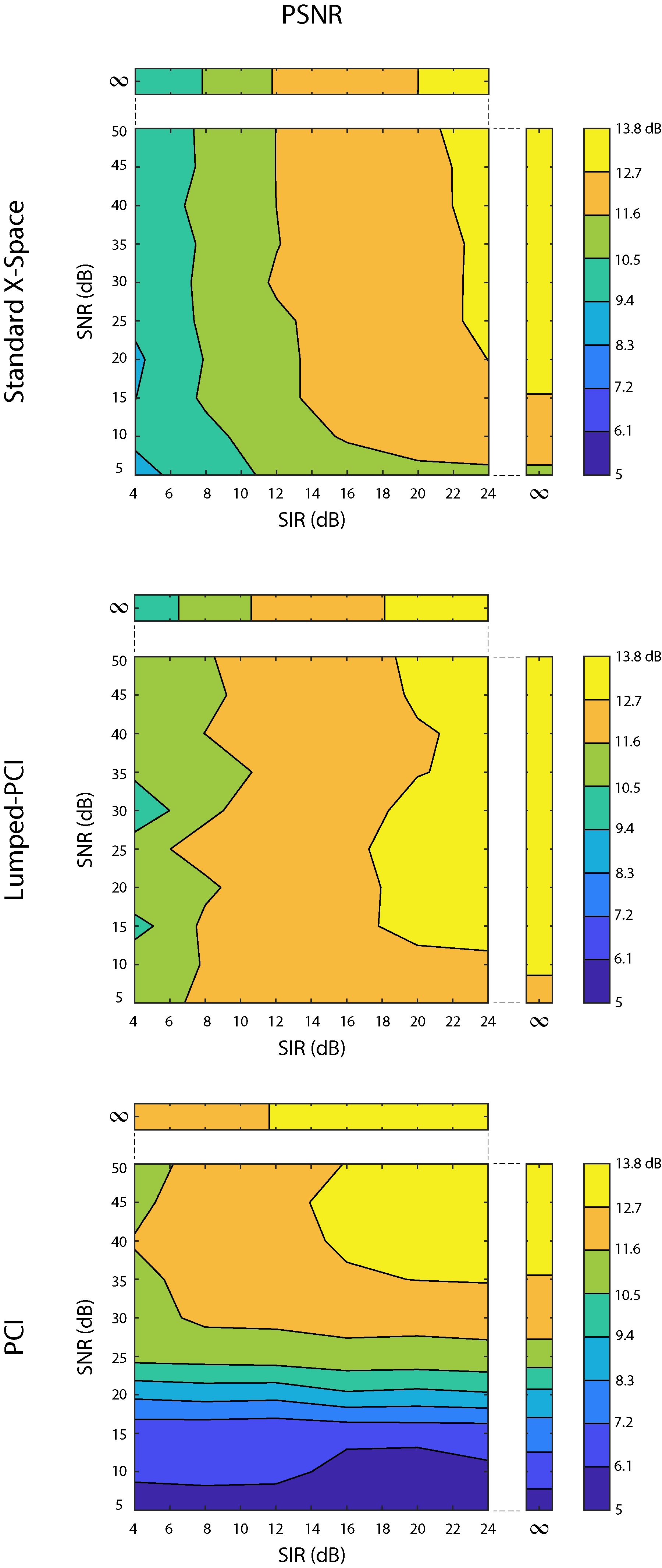} 
\caption{PSNR contour plots showing image quality as a function of SNR and SIR. The individual effects of noise and harmonic interference can be seen at SIR=$\infty$ and SNR=$\infty$ levels, respectively. The performance of standard x-space is mainly dependent on the interference level and not noise, except for SNR $<$ 20 dB. Lumped-PCI outperforms standard x-space at all SNR and SIR levels. PCI yields the highest image quality when interference dominates over noise. PCI outperforms standard x-space when SNR $>$ 25 dB, and Lumped-PCI when SNR $>$ 30 dB.}
\label{fig: psnr2D}
\end{figure}

 Next, the image qualities of the three methods were compared quantitatively using the PSNR metric. At each SNR and SIR level, Monte Carlo simulations were performed by repeating the simulations 50 times, and the resulting PSNR values were averaged across repeats. The contour plots of the mean PSNR values are shown in Fig. \ref{fig: psnr2D}, where the individual effects of noise and harmonic interference are also provided (see SIR=$\infty$ and SNR=$\infty$ levels, respectively). According to this figure, the main factor that determines the performance of standard x-space is the interference level and not noise. Only for SNR $<$ 20 dB, the performance depends jointly on the noise and interference levels. While the performance trends for Lumped-PCI are similar, it outperforms standard x-space at all SNR and SIR levels. The PSNR difference between Lumped-PCI and standard x-space increases as SNR and SIR decreases, e.g., the difference reaches 1.7 dB at SNR = 10 dB and SIR = 4 dB. As expected, with its robustness against interference, PCI outperforms standard x-space at moderate-to-high SNR levels (for SNR $>$ 25 dB). PCI also outperforms Lumped-PCI when SNR $>$ 30 dB. For example, at SNR = 35 dB and SIR = 8 dB, the PSNR values are 10.6 dB, 10.9 dB, and 12.0 dB for standard x-space, Lumped-PCI, and PCI, respectively. At low SNR levels, however, the performance of PCI quickly degrades, as noise effects dominate over interference. 


\par
\begin{figure}[h]
\centering
\includegraphics{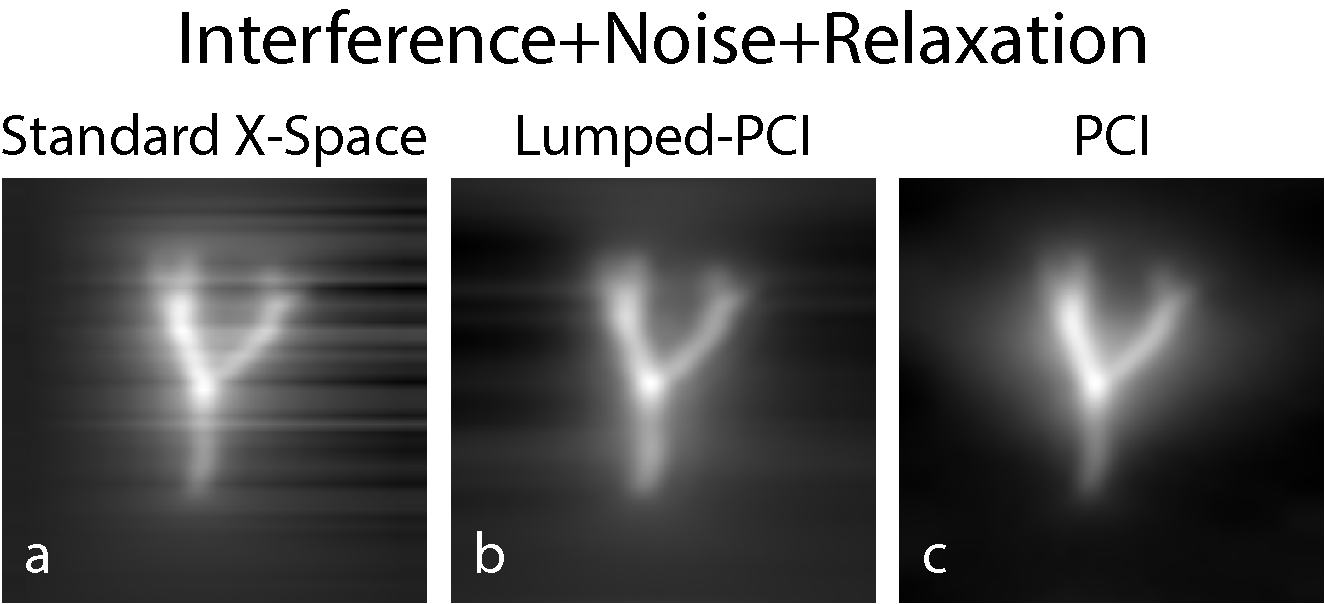}
\caption{Simulation results for a realistic scenario where noise, harmonic interference, and relaxation effects are all incorporated: 30 dB SNR, 8 db SIR, relaxation time constant of $\tau = 3$ $\mu s$. The MPI images reconstructed by (a) standard x-space, (b) Lumped-PCI, and (c) PCI exhibit a slight blurring due to relaxation, but the overall effects of noise and interference remain the same as in Fig. \ref{fig: 88 noise+interference}. Under these conditions, PCI provides the best image quality, whereas Lumped-PCI also shows improved quality when compared to standard x-space.}
\label{fig: 88 relaxation images}
\end{figure}

 Figure \ref{fig: 88 relaxation images} displays the results of the proposed and comparison reconstructions for the combined effects of noise, interference, and relaxation. To investigate the effects of relaxation, we assumed a realistic relaxation time constant of $\tau$ $=$ $3$ $\mu s$ \cite{8322193}. For this analysis, SNR was fixed at 30 dB and SIR at 8 dB. Comparing Fig. \ref{fig: 88 relaxation images} with the third column of Fig. \ref{fig: 88 noise+interference}, one can see that relaxation causes a slight blurring for all three methods. The overall effects of noise and interference, however, remain the same as before. Once again, horizontal stripe artifacts are seen in the standard x-space image. Both Lumped-PCI and PCI provide improved image quality with respect to standard x-space. Under these specific conditions, PCI outperforms the other two methods. The PSNR values corresponding to standard x-space, Lumped-PCI, and PCI are 10.6 dB, 11.9 dB, and 12.6 dB, respectively.

\subsection{Imaging Experiment Results}
\begin{figure}[h]
\centering
\includegraphics{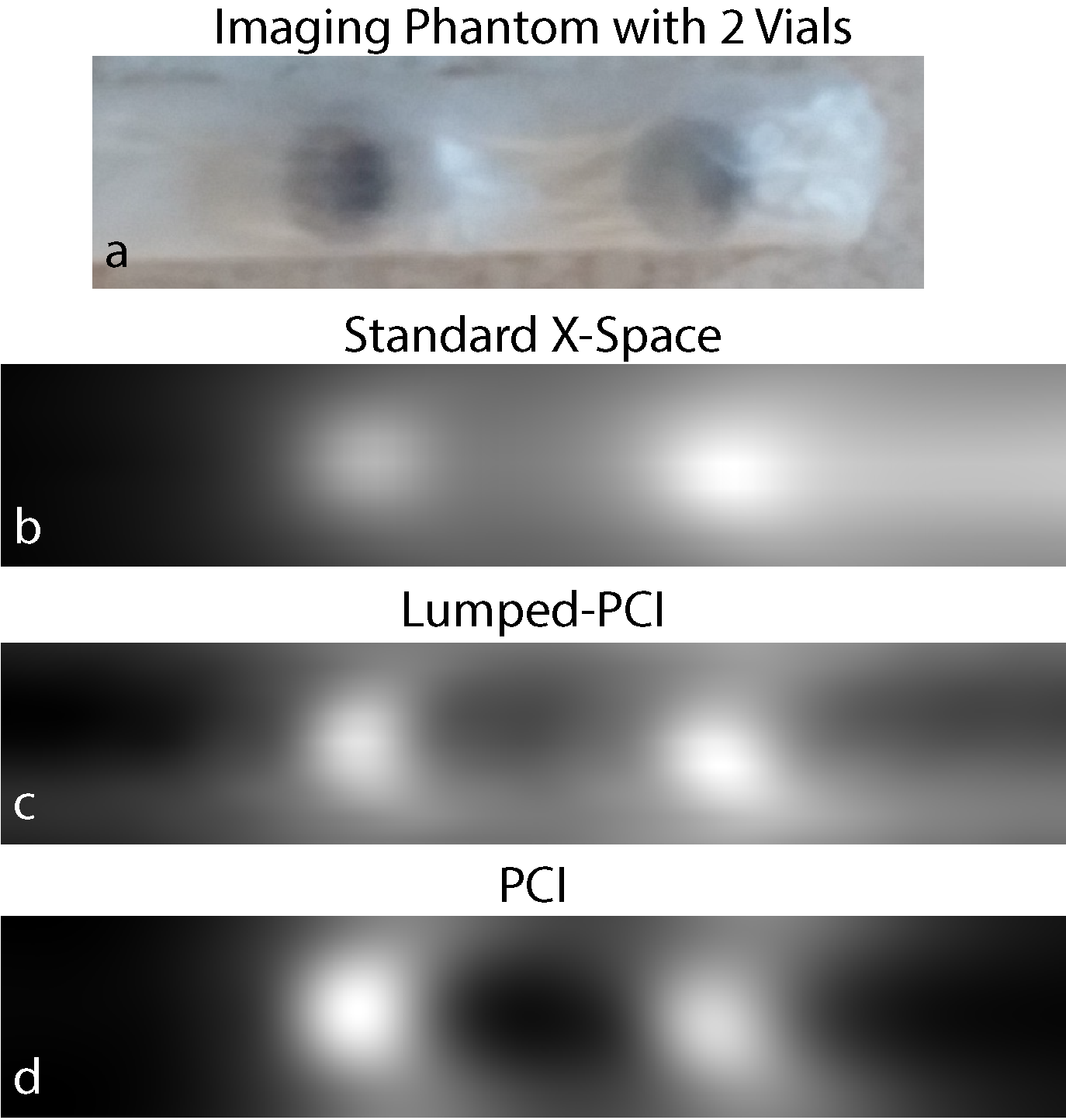}
\caption{Experimental imaging results using (a) an imaging phantom with two vials filled with identical concentration of Perimag nanoparticles, separated at 9-mm distance. The reconstructed MPI images from (b) standard x-space, (c) Lumped-PCI, and (d) PCI. Standard x-space suffers from a pile-up artifact in image intensity due to non-ideal signal conditions, whereas Lumped-PCI provides improved image quality with similar but less severe artifacts along the horizontal direction. PCI does not exhibit any artifacts and provides the highest image quality out of the three methods. FOV size: 0.7$\times$7.52 cm$^2$, displayed FOV size: 0.7$\times$4.7 cm$^2$.}
\label{fig: expPvsP}
\end{figure}

Figure \ref{fig: expPvsP} displays the imaging experiment results of the proposed and comparison reconstructions using a phantom that contains two vials of Perimag nanoparticles separated by a 9-mm distance. Standard x-space suffers from a pile-up artifact in image intensity due to non-ideal signal conditions. Here, imaging experiments utilized 9 lines to cover the 2D FOV, as opposed to 51 lines used in the simulations. Therefore, the visual manifestation of the aforementioned horizontal stripe artifact is now a more dominant pile-up artifact. While Lumped-PCI also demonstrates similar but less severe artifacts along the horizontal direction, the image quality is visibly improved when compared to standard x-space. On the other hand, PCI does not exhibit any artifacts and provides the highest image quality out of the three methods. 

\par The experimental results in Fig. \ref{fig: expN,N+V,V} show that the proposed methods can handle different nanoparticle types. Here, three vials filled with Nanomag-MIP, VivoTrax, and a homogeneous mixture of the two separated by 15-mm distances were imaged. Once again, standard x-space suffers from an intensity pile-up on the right hand side of the image. While the image quality of Lumped-PCI is improved with respect to standard x-space, the effects of interference can still be observed in the horizontal direction. PCI is free from such artifacts and provides the highest image quality due to its robustness against harmonic interference effects. 

\par For the experiments in Fig. \ref{fig: expPvsP} and \ref{fig: expN,N+V,V}, the SNR and SIR levels were computed from the MPI signals in time domain and frequency domain, respectively. These computations yielded SNR = 32.3 dB and SIR = 7.9 dB for the experiments in Fig. \ref{fig: expPvsP}, and SNR = 31.1 dB and SIR = 4.8 dB for those in Fig. \ref{fig: expN,N+V,V}. The parameters for the simulations in Fig. \ref{fig: 88 relaxation images} were based on these experimental results. One can see that the overall effects of non-ideal signal conditions in the experiments are consistent with those seen in the simulations. These proof-of-concept experiments demonstrate both the feasibility with different nanoparticle types and robustness against non-ideal signal conditions for the proposed methods.
\begin{figure}[h]
\centering
\includegraphics{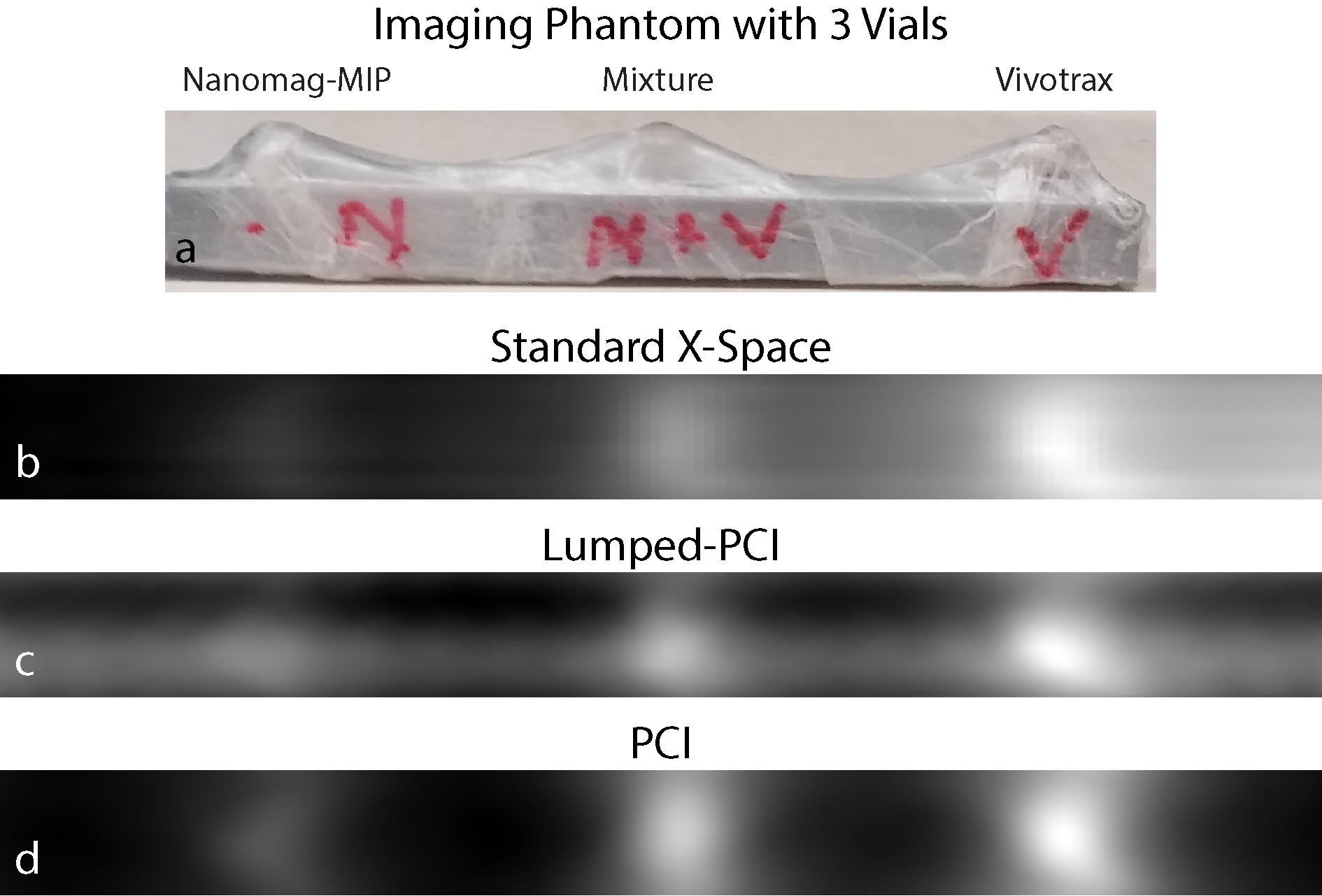}
\caption{Experimental imaging results using different types of nanoparticles. (a) The imaging phantom of three vials filled with Nanomag-MIP, Vivotrax, and a homogeneous mixture of the two. The reconstructed MPI images from (b) standard x-space, (c) Lumped-PCI, and (d) PCI. PCI gives the highest image quality out of the three methods. The results show that the proposed methods can successfully handle different nanoparticle types. FOV size: 0.7$\times$7.52 cm$^2$. }
\label{fig: expN,N+V,V}
\end{figure}

\section{Discussion}
In this work, with simulations and experimental results, we have shown that PCI provides improved robustness against harmonic interferences when compared to standard x-space reconstruction. In cases where using only a small portion of the received signal renders PCI sensitive to noise, we propose to improve its noise robustness by lumping the signals from the entire pFOV. Utilizing the edges of the pFOVs reduces interference robustness of Lumped-PCI, hence, a clear trade-off between noise robustness and interference robustness emerges. To adjust to a specific SNR and SIR level, a central region of pFOV can be lumped instead of the entire pFOV. For example, the signals at pFOV edges can be discarded without significant data loss to improve interference robustness. Similarly, standard x-space reconstruction also proposed using the central 95\% of pFOV to avoid velocity-compensation-induced noise amplification at the edges \cite{6494648}. Note that while this procedure improves noise robustness of x-space reconstruction, it does not provide sufficient robustness against interference. In fact, the experimental results in Fig. \ref{fig: expPvsP}b and Fig. \ref{fig: expN,N+V,V}b utilized the central 90\% of pFOV for x-space reconstruction, but still exhibited pile-up artifacts.
\par There are multiple potential sources of harmonic interferences in MPI, most notably the non-linearities in the transmit chain (e.g., from power amplifier or capacitors). Especially if the active/passive compensation of the direct feedthrough does not provide sufficient decoupling between the transmit and receive coils, the higher harmonics of the drive field can also feed through to the MPI signal. In addition, insufficient shielding may cause eddy currents on the selection field permanent magnets, which in turn can induce signal on the receive coil \cite{Anx-spacemagneticparticleimagingscanner,Dilek}. Likewise, ambient interferences may also become problematic due to insufficient shielding of the receive coil and/or transmit/receive filter chains \cite{EddyshieldedRelaxometer}. Under low signal conditions, such as \textit{in vivo} cases, these harmonic interferences can become a problem even for well-tuned systems. Therefore, the reconstruction technique proposed in this work can improve image quality not just for low-cost MPI scanners (e.g., like the FFP scanner used in this work), but also for high-fidelity commercial systems. Moreover, the proposed PCI method substantially simplifies the reconstruction procedure by eliminating the need for individual pFOV processing (i.e., gridding, DC shift calculation, and stitching). 

\par Another advantage of the proposed method is that, unlike standard x-space reconstruction, it does not require pFOVs to overlap. Hence, the safety limits on the drive field amplitude become less of a problem \cite{6510534}. In fact, the sizes of the kernels in Eqs. \ref{eqn:PCI kernel} and \ref{eqn:PLI kernel} become more compact at smaller drive field amplitudes, which would yield a higher fidelity deconvolution with reduced noise-amplification effects. In return, however, the proposed method requires pFOV centers to be closely spaced. Nevertheless, due to the 20 T/s safety limit on the slew rate of the focus field \cite{slewlimit}, realistic scan trajectories must already consist of closely spaced pFOVs. For the experiments in this work, using a drive field at 9.7 kHz with a slew rate of 70 mT/s yielded 3-$\mu$m distances between pFOV centers. This level of spacing is more than sufficient to reconstruct a high quality image, as it is three orders of magnitude below the expected mm-range resolution of our scanner. When operating at the safety limit of 20 T/s slew rate \cite{slewlimit}, the distances between pFOV centers would scale up to 0.86 mm. While this spacing may also be sufficient in most cases, the slew rate of the focus field can be reduced accordingly to attain a desired spacing level. Demonstration of PCI at such high slew rates remains as future work, as it requires incorporating electromagnetically driven focus fields. 

\section{Conclusion}
In this work, we have proposed a robust x-space image reconstruction that consists of two simplified steps: forming a raw image by directly assigning the signal to pFOV center locations, and deconvolving the raw image by a known, narrow kernel. 
Extensive simulation results and imaging experiments demonstrated that the proposed PCI method outperforms standard x-space reconstruction. PCI is particularly robust against harmonic interferences, making it a promising technique for \textit{in vivo} applications. In SNR starved cases, the noise robustness of PCI can be further improved by lumping the signals from the entire pFOV, with a trade-off of slightly reduced interference robustness. The proposed method promises a computationally simpler and straightforward reconstruction with high-fidelity image reconstruction.

\section*{Acknowledgment}
The authors would like to thank Omer Burak Demirel for his contributions to the MPI simulation toolbox, and Mustafa Utkur for his assistance in the experimental setup. 

\ifCLASSOPTIONcaptionsoff
  \newpage
\fi



%



\bibliographystyle{ieeetr} 
\bibliography{ref}

\begin{thebibliography}{10}

\bibitem{Gleich2005}
B.~Gleich and J.~Weizenecker, ``Tomographic imaging using the nonlinear
  response of magnetic particles,'' {\em Nature}, vol.~435, no.~7046,
  pp.~1214--1217, 2005.

\bibitem{Weizenecker_2007}
J.~Weizenecker, J.~Borgert, and B.~Gleich, ``A simulation study on the
  resolution and sensitivity of magnetic particle imaging,'' {\em Physics in
  Medicine and Biology}, vol.~52, pp.~6363--6374, Oct. 2007.

\bibitem{X-spaceSafe}
P.~W. Goodwill, E.~U. Saritas, L.~R. Croft, T.~N. Kim, K.~M. Krishnan, D.~V.
  Schaffer, and S.~M. Conolly, ``X-space {MPI}: Magnetic nanoparticles for safe
  medical imaging,'' {\em Advanced Materials}, vol.~24, no.~28, pp.~3870--3877,
  2012.

\bibitem{SARITAS2013116}
E.~U. Saritas, P.~W. Goodwill, L.~R. Croft, J.~J. Konkle, K.~Lu, B.~Zheng, and
  S.~M. Conolly, ``Magnetic particle imaging ({MPI}) for {NMR} and {MRI}
  researchers,'' {\em Journal of Magnetic Resonance}, vol.~229, pp.~116 -- 126,
  2013.

\bibitem{Bauer2015}
L.~M. Bauer, S.~F. Situ, M.~A. Griswold, and A.~C.~S. Samia, ``Magnetic
  particle imaging tracers: State-of-the-art and future directions,'' {\em The
  Journal of Physical Chemistry Letters}, vol.~6, pp.~2509--2517, Jul. 2015.

\bibitem{Zheng2017}
B.~Zheng, E.~Yu, R.~Orendorff, K.~Lu, J.~J. Konkle, Z.~W. Tay, D.~Hensley,
  X.~Y. Zhou, P.~Chandrasekharan, E.~U. Saritas, P.~W. Goodwill, J.~D. Hazle,
  and S.~M. Conolly, ``Seeing {SPIOs} directly in vivo with magnetic particle
  imaging,'' {\em Molecular Imaging and Biology}, vol.~19, pp.~385--390, Jun.
  2017.

\bibitem{6510534}
E.~U. {Saritas}, P.~W. {Goodwill}, G.~Z. {Zhang}, and S.~M. {Conolly},
  ``Magnetostimulation limits in magnetic particle imaging,'' {\em IEEE
  Transactions on Medical Imaging}, vol.~32, pp.~1600--1610, Sep. 2013.

\bibitem{schmaleSafety}
I.~{Schmale}, B.~{Gleich}, J.~{Rahmer}, C.~{Bontus}, J.~{Schmidt}, and
  J.~{Borgert}, ``{MPI} safety in the view of {MRI} safety standards,'' {\em
  IEEE Transactions on Magnetics}, vol.~51, pp.~1--4, Feb. 2015.

\bibitem{6494648}
K.~{Lu}, P.~W. {Goodwill}, E.~U. {Saritas}, B.~{Zheng}, and S.~M. {Conolly},
  ``Linearity and shift invariance for quantitative magnetic particle
  imaging,'' {\em IEEE Transactions on Medical Imaging}, vol.~32,
  pp.~1565--1575, Sep. 2013.

\bibitem{patch1}
M.~Gruttner, T.~F. Sattel, F.~Griese, and T.~M. Buzug, ``{System matrices for
  field of view patches in magnetic particle imaging},'' in {\em Medical
  Imaging 2013: Biomedical Applications in Molecular, Structural, and
  Functional Imaging} (J.~B. Weaver and R.~C. Molthen, eds.), vol.~8672,
  pp.~305 -- 310, International Society for Optics and Photonics, SPIE, 2013.

\bibitem{patch2}
P.~{Szwargulski}, M.~{Ahlborg}, C.~{Kaethner}, and T.~M. {Buzug}, ``Trajectory
  analysis using static patches for magnetic particle imaging,'' {\em IEEE
  Transactions on Magnetics}, vol.~51, pp.~1--4, Feb. 2015.

\bibitem{patch3}
T.~{Knopp} and M.~{Kaul}, ``{MPI} focus field experiments using non-overlapping
  focus-field patches,'' in {\em Proc. 5th Int. Workshop Magn. Part. Imag.
  (IWMPI), Istanbul, Turkey}, Mar. 2015.

\bibitem{slew-rate}
J.~J. {Konkle}, P.~W. {Goodwill}, E.~U. {Saritas}, B.~{Zheng}, K.~{Lu}, and
  S.~M. {Conolly}, ``Twenty-fold acceleration of 3d projection reconstruction
  mpi,'' {\em Biomedizinische Technik. Biomedical engineering}, vol.~58,
  pp.~1--12, Aug. 2013.

\bibitem{slewlimit}
{International Commission on Non-Ionizing Radiation Protection (ICNIRP)},
  ``{{M}edical magnetic resonance ({M}{R}) procedures: protection of
  patients},'' {\em Health Phys}, vol.~87, pp.~197--216, Aug. 2004.

\bibitem{5482192}
P.~W. {Goodwill} and S.~M. {Conolly}, ``The x-space formulation of the magnetic
  particle imaging process: 1-d signal, resolution, bandwidth, snr, sar, and
  magnetostimulation,'' {\em IEEE Transactions on Medical Imaging}, vol.~29,
  pp.~1851--1859, Nov. 2010.

\bibitem{5728922}
P.~W. {Goodwill} and S.~M. {Conolly}, ``Multidimensional x-space magnetic
  particle imaging,'' {\em IEEE Transactions on Medical Imaging}, vol.~30,
  pp.~1581--1590, Sep. 2011.

\bibitem{convex_x_space}
J.~J. Konkle, P.~W. Goodwill, D.~W. Hensley, R.~D. Orendorff, M.~Lustig, and
  S.~M. Conolly, ``A convex formulation for magnetic particle imaging x-space
  reconstruction,'' {\em PLOS ONE}, vol.~10, pp.~1--15, Oct. 2015.

\bibitem{alper}
A.~A. Ozaslan, A.~Alacaoglu, O.~B. Demirel, T.~{\c{C}}ukur, and E.~U. Saritas,
  ``Fully automated gridding reconstruction for non-cartesian x-space magnetic
  particle imaging,'' {\em Physics in Medicine {\&} Biology}, vol.~64,
  p.~165018, Aug. 2019.

\bibitem{pulsedMPI}
Z.~W. {Tay}, D.~{Hensley}, J.~{Ma}, P.~{Chandrasekharan}, B.~{Zheng},
  P.~{Goodwill}, and S.~{Conolly}, ``Pulsed excitation in magnetic particle
  imaging,'' {\em IEEE Transactions on Medical Imaging}, vol.~38,
  pp.~2389--2399, Oct. 2019.

\bibitem{7107082}
M.~{Utkur} and E.~U. {Saritas}, ``Comparison of different coil topologies for
  an mpi relaxometer,'' in {\em Proc. 5th Int. Workshop Magn. Part. Imag.
  (IWMPI), Istanbul, Turkey}, Mar. 2015.

\bibitem{ProjectionXspace}
P.~W. {Goodwill}, J.~J. {Konkle}, B.~{Zheng}, E.~U. {Saritas}, and S.~M.
  {Conolly}, ``Projection x-space magnetic particle imaging,'' {\em IEEE
  Transactions on Medical Imaging}, vol.~31, pp.~1076--1085, May 2012.

\bibitem{Analogreceivesignalprocessing}
M.~Graeser, T.~Knopp, M.~GrÃ¼ttner, T.~F. Sattel, and T.~M. Buzug, ``Analog
  receive signal processing for magnetic particle imaging,'' {\em Medical
  Physics}, vol.~40, no.~4, p.~042303, 2013.

\bibitem{ActivePassiveCompensation}
D.~Pantke, N.~Holle, A.~Mogarkar, M.~Straub, and V.~Schulz, ``Multifrequency
  magnetic particle imaging enabled by a combined passive and active drive
  field feed-through compensation approach,'' {\em Medical Physics}, vol.~46,
  no.~9, pp.~4077--4086, 2019.

\bibitem{gazimmfd337864}
E.~Bozkurt and E.~U. Saritas, ``Signal-to-noise ratio optimized image
  reconstruction technique for magnetic particle imaging,'' {\em J. Fac. Eng.
  Archit. Gaz.}, vol.~32, pp.~999--1013, 2017.

\bibitem{rahmersignalencoding}
J.~Rahmer, J.~Weizenecker, B.~Gleich, and J.~Borgert, ``Signal encoding in
  magnetic particle imaging: Properties of the system function,'' {\em BMC
  medical imaging}, vol.~9, p.~4, May 2009.

\bibitem{7960614}
D.~{Sarica}, O.~B. {Demirel}, and E.~U. {Saritas}, ``{DC} shift based image
  reconstruction for magnetic particle imaging,'' in {\em Proc. 25th Signal
  Processing and Communications Applications Conference (SIU)}, May 2017.

\bibitem{HDX-IWMPI}
S.~{Kurt}, Y.~{Muslu}, M.~{Utkur}, and E.~U. {Saritas}, ``Harmonic dispersion
  x-space {MPI},'' in {\em Proc. 9th Int. Workshop Magn. Part. Imag. (IWMPI),
  New York, NY, USA}, pp.~75--76, Mar. 2019.

\bibitem{8322193}
Y.~{Muslu}, M.~{Utkur}, O.~B. {Demirel}, and E.~U. {Saritas},
  ``Calibration-free relaxation-based multi-color magnetic particle imaging,''
  {\em IEEE Transactions on Medical Imaging}, vol.~37, pp.~1920--1931, Aug.
  2018.

\bibitem{mustafaviscomakale}
M.~Utkur, Y.~Muslu, and E.~U. Saritas, ``Relaxation-based color magnetic
  particle imaging for viscosity mapping,'' {\em Applied Physics Letters},
  vol.~115, no.~15, p.~152403, 2019.

\bibitem{6297476}
L.~R. {Croft}, P.~W. {Goodwill}, and S.~M. {Conolly}, ``Relaxation in x-space
  magnetic particle imaging,'' {\em IEEE Transactions on Medical Imaging},
  vol.~31, pp.~2335--2342, Dec. 2012.

\bibitem{8479214}
M.~{Utkur}, Y.~{Muslu}, and E.~U. {Saritas}, ``A 4.8 {T}/m magnetic particle
  imaging scanner design and construction,'' in {\em Proc. 21st National
  Biomedical Engineering Meeting (BIYOMUT), Istanbul, Turkey}, Nov. 2017.

\bibitem{Anx-spacemagneticparticleimagingscanner}
P.~W. Goodwill, K.~Lu, B.~Zheng, and S.~M. Conolly, ``An x-space magnetic
  particle imaging scanner,'' {\em Review of Scientific Instruments}, vol.~83,
  no.~3, p.~033708, 2012.

\bibitem{Dilek}
D.~M. {Yalcinkaya}, M.~{Utkur}, and E.~U. {Saritas}, ``Finite element analysis
  of passive magnetic shields for a {FFP} {MPI} scanner,'' in {\em Proc. 8th
  Int. Workshop Magn. Part. Imag. (IWMPI), Hamburg, Germany}, pp.~123--124,
  Mar. 2018.

\bibitem{EddyshieldedRelaxometer}
L.~M. Bauer, D.~W. Hensley, B.~Zheng, Z.~W. Tay, P.~W. Goodwill, M.~A.
  Griswold, and S.~M. Conolly, ``Eddy current-shielded x-space relaxometer for
  sensitive magnetic nanoparticle characterization,'' {\em Review of Scientific
  Instruments}, vol.~87, no.~5, p.~055109, 2016.

\end{thebibliography}
%






\end{document}